\documentclass[conference]{IEEEtran}
\IEEEoverridecommandlockouts
\usepackage{cite}
\usepackage[english]{babel}

\usepackage[T1]{fontenc}
\usepackage{multirow}
\usepackage{amsmath,amssymb,amsfonts}
\usepackage{multicol}
\usepackage{multirow}
\usepackage{graphicx}
\usepackage{subfigure}
\usepackage{color}
\usepackage{textcomp}
\usepackage{soul}
\usepackage{tagging}

\droptag{Atal_personal}

\usepackage{array}
\newcolumntype{L}{>{\centering\arraybackslash}m{3cm}}

\sethlcolor{yellow}
\usepackage[colorinlistoftodos,prependcaption,textsize=normalsize]{todonotes}
\usepackage{xargs}

\newcommandx{\doubt}[2][1=]{\todo[linecolor=red,backgroundcolor=red!25,bordercolor=red,#1]{#2}}

\usepackage[colorlinks=true, allcolors=blue]{hyperref}
\def\BibTeX{{\rm B\kern-.05em{\sc i\kern-.025em b}\kern-.08em
        T\kern-.1667em\lower.7ex\hbox{E}\kern-.125emX}}
\newcommand*{\email}[1]{%
    \normalsize\href{mailto:#1}{#1}\par
    }
\begin{document}
    
    \title{Automated Crater Detection from Co-registered Optical Images, Elevation Maps and Slope Maps using Deep Learning
        
    }
    
    %
    \author{
    \IEEEauthorblockN{Atal Tewari\IEEEauthorrefmark{2}, Vinay Verma\IEEEauthorrefmark{2}, Pradeep Srivastava\IEEEauthorrefmark{3}, Vikrant Jain\IEEEauthorrefmark{3}, Nitin Khanna\IEEEauthorrefmark{2}$^{,\star}$\thanks{$^{\star}$ Corresponding author. Please address all correspondences to Nitin Khanna, Multimedia Analysis and Security (MANAS) Lab, Electrical Engineering, Indian Institute of Technology Gandhinagar, India. E-mail address: \email{nitinkhanna@iitgn.ac.in}}}
    
    \IEEEauthorblockA{\IEEEauthorrefmark{2}Electrical Engineering, Indian Institute of Technology Gandhinagar, India}
    \IEEEauthorblockA{\IEEEauthorrefmark{3} Earth Sciences, Indian Institute of Technology Gandhinagar, India}
}

        
    
\maketitle  
\thispagestyle{plain}
\pagestyle{plain}

\begin{abstract}
Impact craters are formed as a result of continuous impacts on the surface of planetary bodies. 
This paper proposes a novel way of simultaneously utilizing optical images, digital elevation maps (DEMs), and slope maps for automatic crater detection on the lunar surface. 
Mask R-CNN, tuned for the crater detection task, is utilized in this paper.
Two catalogs, namely, Head-LROC and Robbins, are used for the performance evaluation.
Exhaustive analysis of the detection results on the lunar surface has been performed with respect to both Head-LROC and Robbins catalog.
With the Head-LROC catalog, which has relatively strict crater markings and larger possibility of missing craters, recall value of 94.28\% has been obtained as compared to 88.03\% for the baseline method.
However, with respect to a manually marked exhaustive crater catalog based on relatively liberal marking, significant precision and recall values are obtained for different crater size ranges.
The generalization capability of the proposed method in terms of crater detection on a different terrain with different input data type is also evaluated.
We show that the proposed model trained on the lunar surface with optical images, DEMs and corresponding slope maps can be used to detect craters on the Martian surface even with entirely different input data type, such as thermal IR images from the Martian surface.

\end{abstract}

\section{Introduction}

The lunar surface has an abundance of craters preserved due to the absence of the atmosphere. 
Moreover, craters are among the most prominent entity of the lunar surface and provide vital information about the surface and its evolution. 
For a long time, crater density has been correlated with surface age, and crater count has been used to estimate the relative age of the lunar surface~\cite{basilevskii1976evolution,michael2010planetary}. 
Alongside, craters can also reveal information about the direction of impact and impactor population~\cite{head2010global}. 
These lunar craters degrade under the effects of gravity and surface processes. Therefore, the study of craters can be used to obtain information about the surface properties and various processes that act on them. 
Further, lunar regolith is created and evolved due to energy generated from the impacts and is related to the sizes and number of impacts.
Hence, the study of lunar craters can also serve as a tool to understand the nature of lunar regolith~\cite{bart2011global}. 

The first step, however, in any such study is to detect and locate craters. 
This step can be addressed either via manual marking or by applying various image processing techniques on contour maps, optical images, digital elevation maps (DEMs), or possible combinations of these three inputs. 
Moreover, automatically mapping craters on the lunar surface, or any other planetary surface, for that matter, can reduce the manual effort. Crater mapping is needed to determine the landing sites and routing planetary rovers based on the in situ data in the absence of GPS, without relying on the communication with Earth stations. 

The lunar surface has varying crater distribution (heavily cratered on highlands and less cratered on maria), sizes (from a few hundred meters to hundreds of kilometers in diameter), shapes (circular, elliptical, concentric, and overlapping), floor structures (flat-floored, round-floored, floor with terraces, floor with central mound), and ages (fresh, moderately degraded, degraded). 
This diversity in crater property, crater quality, and the availability of different input data types such as DEM and optical image, pose several unique challenges for crater detection.

Many researchers have proposed crater detection systems that utilize both optical images and DEMs.  
All such systems can be broadly divided into three categories. 
The first category consists of systems that utilize optical image for detecting the craters and DEM for verification purposes only~\cite{degirmenci2010impact, kang2015automatic}.  
The second category consists of systems that utilize optical images and DEMs as independent information sources~\cite{salamuniccar2014integrated}. 
The second category systems process both optical image and DEM independently and then combine their corresponding results to generate the final crater map. 
The most recent systems of this category are proposed in ~\cite{yang2019bayesian, zuo2019shadow}.  
In~\cite{yang2019bayesian}, features are extracted from DEM and optical image independently and then separately used to train two support vector machine (SVM) models. 
These SVMs give the intermediate class posterior probability values. Finally, the Bayesian-based network combines all the intermediate class posterior probability values and obtains the final detection. In~\cite{zuo2019shadow}, small size craters ([1, 10] km diameter) were detected, which utilized the shadow and highlight features from optical images.  
For medium to large size craters ([5, 300] km diameter) DEM based method~\cite{zuo2016contour} was used. 
Finally, the system integrated the results from both approaches.
The third category utilizes optical image and DEM to generate more training samples~\cite{wang2019active}. 
In contrast to the above categories, this paper proposes a system that performs fusion at the input stage and then extracts information, while earlier papers performed decision fusion at later stages of information extraction. 
The proposed deep learning-based system for crater detection utilizes optical image and DEM data simultaneously. 
We first combine the optical image and DEM data and then pass them through a deep learning framework, i.e., Mask R-CNN for crater detection. 
This early fusion of data and usage of a deep learning framework makes the proposed system more robust. It allows the system to detect craters on any given region of planetary bodies, regardless of the underlying terrain.

The main contributions of this paper are as follows:
\begin{itemize}
    \item Optical images and DEMs and corresponding slope have been used simultaneously for crater detection without extracting any hand-crafted features. 
	\item Crater detection is done on two human-generated catalogs~\cite{head2010global},~\cite{povilaitis2018crater} combined, and~\cite{robbins2019new}. Localization performance is also evaluated. 
	\item For the first time, Robbins~\cite{robbins2019new} lunar catalog (contains the largest number of manually marked craters) is used for training crater detection models.
	\item We explore the algorithm's generalization capability on a different planetary surface, i.e., Mars. A new input data type independent crater detection field is explored, i.e., training on the concatenated input of optical image, DEM, and corresponding slope while detecting on near IR images.

\end{itemize}

The rest of this paper is organized as follows. 
Section~\ref{sec:related_works} describes the existing literature in the field.
The proposed system and its components are detailed in Section~\ref{sec:methods}. 
Further, Section~\ref{sec:experimental results} presents a detailed analysis of the experimental results. 
Conclusions and future work are presented in Section~\ref{sec:Conclusion and Future Work}.

\section{Related Works}
\label{sec:related_works}

Many studies were carried out on manually marked craters in the early 20th century~\cite{pike1976crater,ravine1986analysis}. 
Manual marking is a tedious task and is prone to errors, with differences existing between markings done by different experts~\cite{bugiolacchi2016moon, robbins2014variability, kirchoff2011examining, greeley1970precision}. In~\cite{robbins2014variability}, it was shown that up to $\sim$ $\pm 45$\% differences exists in the manually marked craters by different experts, which depends on diameter, total craters per diameter bin, and terrain type. 
Nevertheless, due to the recent advances in satellite imaging technology, an ample amount of high-resolution data has become available. 
This high-resolution data enables us to detect numerous small size craters and estimate more refined features of large craters.
The lunar surface harbors numerous craters of varied sizes, shapes, and types, and the number of craters rapidly increases with a decrease in crater size. 
Therefore, accurate manual detection of all these craters is not feasible.  
Thus, several studies have focused on designing algorithms for automatic crater detection from high-resolution images without manual intervention. 
In the following sections, we present the existing literature on crater detection on the planetary surfaces. First, we broadly categorized the existing literature based on the input data type (optical image and DEM). 
Finally, recent methods based on the deep learning framework are presented in Section~\ref{subsec:rel_deep}.

\subsection{Crater Detection using optical images}

In optical images, craters can be visualized as circular features with distinct bright and shadow patterns and are differentiated from the background by crater rim~\cite{vijayan2013crater, bandeira2012detection}. 

One common approach towards crater detection using optical images, involves performing edge detection and then using template matching or Hough transform to retrieve final craters~\cite{kim2005automated}. 
Furthermore, there was a sufficient number of false detected craters that can be removed by various strategies such as the use of eigencraters and neural networks, as described in~\cite{kim2005automated} and thresholding of circularity index used in~\cite{sawabe2005automated}.
Kang et al.~\cite{kang2015automatic} derived the features from camera images and DEM data to identify craters.

In~\cite{bandeira2007impact}, authors first binarized a given image and further used template matching to construct a probability volume.
A 2-D image defined the first two dimensions of the probability volume, and the radii of template craters defined the third dimension. 
The voxel with the highest probability value would give the location and the radius of the impact crater.  
Authors in~\cite{martins2009crater} used different masks to extract Haar-like features from the images. 
An AdaBoost based approach was used to extract useful features from this set and subsequently classify each region into a crater or non-crater. 

The bright and shadow patterns for detecting craters with the optical images is one of the commonly used techniques. 
One such study used thresholding to identify the bright and shadow parts of the craters, which were paired up based on their respective sizes and distance between them~\cite{vijayan2013crater}. 
That resulted in removing those areas that could not be properly matched with their corresponding bright or shadow part. 
Another study~\cite{bandeira2010automatic} used shape filters first to detect the crater candidates, and further calculated texture features for the resulted crater candidates, and finally trained the AdaBoost classifier to identify the craters. 

\subsection{Crater Detection using Digital Elevation Maps (DEMs)}


In DEM, the craters are visualized as circular depressions with closed concentric contours in the valley part.
A flooding algorithm was proposed to detect the depressions in the study~\cite{o1984extraction}. 
However, the problem with the flooding algorithm was that it could not differentiate between two craters, which were superimposed or largely intersecting~\cite{bue2007machine}. 
An improvement in this algorithm was suggested by~\cite{stepinski2009machine}, which introduces a rule-based system, AutoCrat, to identify topographic depressions. 
To overcome the problem of superimposed craters, the authors identified craters in ascending order of their size. 
They also used decision trees to improve the accuracy of crater detection. 
In another study, various shape filters were used to remove non-circular closed contours to generate a final set of candidate craters~\cite{yue2013shape}. 
The authors of the study analyzed the concavity of these candidate craters to identify true craters. \\
Many studies also utilize the slope map derived from DEM to detect craters. A novel algorithm, rotational pixel swapping, was designed by a group of researchers~\cite{yamamoto2017automated} to detect the craters by exploiting the circularity of craters. 
To reduce the search space, they used the heuristics that the crater slope lies in the range of $10^{\circ}$-$33^{\circ}$. 
In one study, terrain attributes like slope and profile curvature were used to select the candidate craters from DEM~\cite{luo2013global}. 
The authors generated a binary image from these terrain attributes with the help of thresholding. 
On the proposed candidate craters, they performed morphological processing and Hough transform to confirm the presence of a true crater. Another research involved a high rate of change of slope of aspect (SOA) to detect the candidate crater rims. 
Furthermore, morphological processing such as thinning, edge linking, noise removal are used to get the true crater rims~\cite{zhou2018automatic}. 
In another study~\cite{chen2018lunar}, authors classified the lunar craters into various types such as con-craters, dispersed craters, connective impact craters, and used different detection algorithms for each type.

Another study~\cite{wang2019active} generated the optimized training data from optical image as well as DEM and used an SVM classifier to detect a crater or non-crater. 

\subsection{Deep Learning Based Approaches}
\label{subsec:rel_deep}

Advancements in big data and deep learning technology improved the state-of-the-art performance in many computer vision-related tasks, especially object detection. 
The advances in GPU technology reduced the time required to converge the deeper and larger networks leading to the development of various CNN architectures, which improved the accuracy of classification tasks. Many researchers have attempted to solve the crater detection problem using all these advances in deep learning and neural networks. 

Performance of CNN and support vector machine (SVM) was compared in a study that detects impact craters, volcanic rootless cones (VRCs), and transverse aeolian ridges (TARs)~\cite{palafox2017automated}. 
Their CNN model comprised of 5 networks trained to detect the objects as mentioned earlier of different sizes. 
For SVM, they extracted histograms of oriented gradients and used them as input to detect the impact craters, VRCs, and TARs. 
Their work proved that CNN performed better as compared to SVM with lesser false positives. 
Another group of researchers~\cite{emami2015automatic} used edge detection and convex grouping to find the candidate crater regions and later used CNN for the final crater and non-crater class~\cite{emami2015automatic}. 
Faster-RCNN,  a popular object detection algorithm, was used to detect the terrain features such as craters, mountains, and geysers~\cite{li2017recognizing}. 
The dataset used in the study~\cite{li2017recognizing} comprised $400$ manually marked images of such terrain features. 
Although the detection algorithm's performance was on the higher side for the study~\cite{li2017recognizing}, it was difficult to comment on their network's generalization capability due to a relatively small dataset used for the performance evaluation. 
Few recent deep learning based studies~\cite{delatte2019segmentation,lee2019automated,silburt2019lunar} used the U-Net architecture to generate crater ring masks. 
Finally, these generated masks were used in template matching to extract crater size and location.
 
Four unsupervised algorithms, namely, convex grouping, hough transform, highlight-shadow region detection, interest points, were used to generate a candidate crater~\cite{emami2019crater}.
Furthermore, the candidate craters were processed with a CNN, having two convolution layers and fully connected layers, to detect the final craters. 
In a recent work~\cite{ali2020automated}, crater detection was done using Mask R-CNN with 87\%  recall on~\cite{head2010global},~\cite{povilaitis2018crater} combined catalogs.
The segmented results from the Mask R-CNN were used for examining the ellipticity distribution of the craters. 
Alongside, the morphological parameters such as depth, depth-diameter ratio were also analyzed. 


\section{Methodology} 
\label{sec:methods}

\begin{figure*}[htb!]
	\includegraphics[width=1\linewidth]{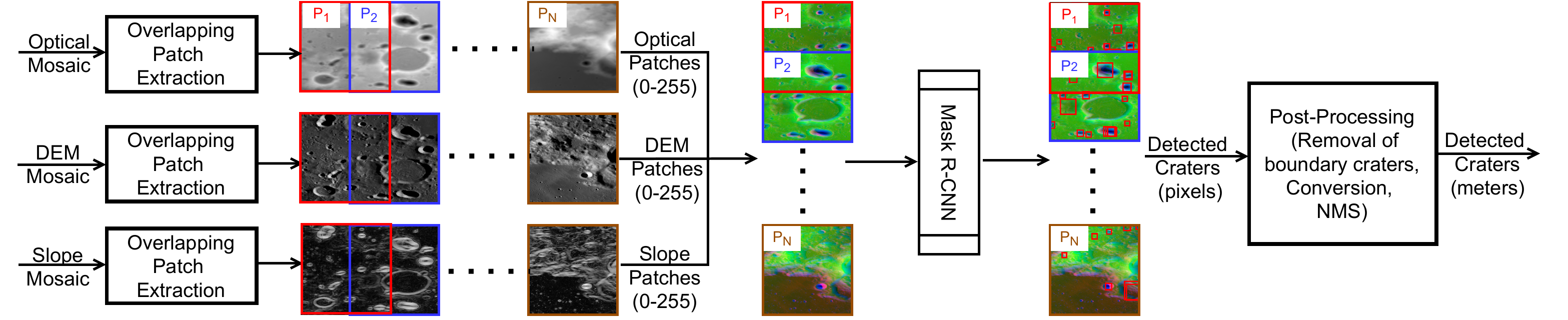}
	\caption{Overview of the Proposed System. 
	Given the mosaic of DEM, optical image, and slope, first, we extract the overlapping patches from them.  
    Further, three-channel patches are formed using the obtained overlapping DEM, optical image, and slope patches.
    These overlapping three-channel patches are passed to the Mask R-CNN to detect the craters. 
    Finally, a post-processing step is applied to the detected craters to remove the duplicate/partial crater and extract the global location. }
	\label{fig:methodology}
\end{figure*}

Crater detection can be thought of as an object detection problem that consists of only one class of objects, namely craters. 
Conventional object detection systems mostly use RGB images where objects are distinguished based on properties such as color, shape, intensity, shadow, and their associations~\cite{zhang2020bridging}.
On the other hand, craters are identified based on a different set of properties, such as bright-shadow patterns, elevation profiles, and slope values~\cite{vijayan2013crater, yamamoto2017automated, luo2013global}. 
The system introduced in this paper reduces the manual effort required in mapping craters by developing an automatic crater detection algorithm. 
For simplicity, this work treats craters as bowl-shaped depressions and does not focus on categorizing them further, such as round or flat-floored craters and primary or secondary craters~\cite{liu2019machine, agarwal2019study}. Figure~\ref{fig:methodology}, gives an overview of our proposed methodology. 
Given the mosaic of DEM, optical image, and slope, first, we extract the overlapping patches from them.  
Further, three-channel patches are formed using the obtained overlapping DEM, optical image, and slope patches.
These overlapping three-channel patches are passed to the Mask R-CNN to detect the craters. 
Finally, a post-processing step is applied to the detected craters to eliminate the duplicate/partial crater and extract the global location. 
A detailed explanation of each step of the proposed framework is described in the following sections.

\subsection{Input to the Network}
\label{sec:region_proposal}
Our proposed algorithm's input is the overlapping three-channel patches obtained from optical, DEM, and slope patches. 
In optical images, the crater's visibility is affected by illumination conditions due to the sun angle. 
While DEMs are not affected by the illumination conditions, but DEM's has the limitation that it lacks context information.  
Hence, using only DEMs for crater detection is likely to miss many shallow craters. 
Apart from the optical images and DEMs, we also utilized the slope raster obtained from the DEMs.
Slope raster has been previously explored for crater detection~\cite{yamamoto2017automated, luo2013global}. 
Craters' slope values approximately lie within a certain range, $10^\circ-33^\circ$~\cite{yamamoto2017automated}.  

Hence, each type of data (optical, DEM, and slope) has its pros and cons. Moreover, using it simultaneously is likely to help deep learning systems to learn useful representations from the combination of three.
The rationale behind the utilization of a combination of optical images and DEMs, is described as follows.
Only one type of input (optical image, DEM) data is available in some planetary surfaces or regions.
In these cases system designed for the optical images is not likely to work with the DEM input data and vice-versa.
While in the proposed approach, although our model is trained with the optical image as well as DEM, at the test time, our method performs well even with the single type of input (Section~\ref{sec:testing_on_martian_thermal_IR_and_DEM_data}) with slightly degraded performances.

We have used the intensity $(I)$, elevation $(E)$, and slope values $(S)$ for the crater detection in the proposed approach. 
Consequently, each feature point is described as $(x,y,i,e,s)$ where $(i,e,s)$ corresponds to the values of intensity, elevation, and slope for the pixel at coordinate $(x, y)$ in the respective images. 
To obtain such point coordinates, we have constructed 3-channel images by providing the optical (intensity) images in the first channel, DEM (elevation) in the second channel, and slope (gradient of elevation) map in the third channel. 
Intuitively for our images, high values in the first channel represent bright parts of craters and ejecta near crater rims. 
Similarly, high values in the second channels represent the high elevation of crater rims and the surrounding region.
Furthermore, the high values in the third channel correspond to the high slope of the crater walls; in other words, it indicates the abrupt change in corresponding adjacent pixels of the second channel. Optical image, DEM, and slope patches are scaled linearly between 0-255 before using them as input to the Mask R-CNN.


\subsection{Mask R-CNN}

Mask R-CNN~\cite{he2017mask} is an instance segmentation method, a combination of semantic segmentation and object detection. 
Semantic segmentation classifies each pixel to a particular class without differentiating between the instances of the same class. 
Object detection is used for classifying and localizing the different instances of objects. 
Hence for a given input image, crater patches in our case, Mask R-CNN provides classification, localization, and masked (segmented) output for each instance of a crater. 
For the crater detection problem tackled in this paper, we have only utilized Mask R-CNN's detection framework.

Mask R-CNN mainly consists of three steps. 
First, it extracts the features using convolution neural networks (such as ResNet~\cite{he2016deep} or ResNeXt~\cite{xie2017aggregated}) with feature pyramid network (FPN)~\cite{lin2017feature} architecture. 
Second, it generates the region proposal from extracted features using the region proposal network (RPN), where the region proposal is a bounding box around a potential crater. 
Finally, the mask prediction is made in parallel with the classification and localization steps.

In our crater detection framework, we have utilized ResNet~\cite{he2016deep}, which is a popular and widely used CNN  architecture. 
As the number of layers increases in a convolutional neural network (CNN) architecture, the training accuracy gets saturated.
To handle this issue, the insight is to copy the shallower network to the deeper network with identity mapping so that the deeper layers have no more training error than the shallower network. 
However, identity mapping is not easy to learn, and the residual learning framework was proposed in~\cite{he2016deep} to tackle the issue, where identity mapping is done by shortcut connections. 
CNN architectures of various depth were proposed in~\cite{he2016deep}. For this paper, we utilized a $50$ layer architecture named ResNet-50 with multiple bottleneck layers. 
The bottleneck layer, 
consists of a sequence of three convolutional layers, starting with the filter size of 1$\times$1, 3$\times$3, and 1$\times$1. 
The number of filters used for these three convolutional layers varies based on the location of the bottleneck layer in the overall ResNet architecture. 
ResNet-50 consists of a total of 5 stages. 
The first stage has one 7$\times$7 convolutional layers with 64 filters.
The second stage consists of three bottleneck layers, where the number of filters for 1$\times$1, 3$\times$3, and 1$\times$1 convolutions are 64, 64, and 128, respectively.
Similarly, the third, fourth, and fifth stages have 4, 6, and 3 bottleneck layers, with the number of filters of size 1$\times$1, 3$\times$3, and 1$\times$1 being double after each stage. 
For example, in the third stage, all the four bottleneck layers will have  128, 128, and 256 filters of size 1$\times$1, 3$\times$3, and 1$\times$1, respectively. 
Furthermore, after each of the five stages, the feature map size is reduced to half.

In hand-engineered features, the feature image pyramid is used to make the algorithm scale-invariant. However, the use of an image pyramid for deep neural networks is infeasible as it increases the time-space complexity. Thus FPN uses the in-network feature hierarchy, which produces the feature on multiple spatial resolutions. FPN tries to combine the high-resolution features with high semantic features. First, a bottom-up pathway is used, which takes a high-resolution image to more semantic features; then, a top-down pathway is used with a lateral connection to combine semantically high features and high-resolution features.

In the past, object detectors generated the candidate region proposals, bounding boxes around potential objects, using traditional algorithms such as a selective search~\cite{uijlings2013selective}, and edge boxes~\cite{zitnick2014edge}. 
Then these region proposals are passed through a deep learning framework to detect the object. 
As region proposal algorithms are implemented in the CPU, in contrast, the deep learning algorithm implemented in GPU makes runtime comparison unfair. 
Also, the runtime is dominated by region proposal generation. 
To overcome this barrier, the RPN is proposed in~\cite{ren2017faster}. 
In which fully convolutional neural networks are used to generate the region proposals.

\subsection{Post Processing}
\label{sec:Post_processing}

We further refine the crater detection output from the Mask R-CNN by applying the following post-processing steps.

\subsubsection{Removing Boundary Craters from Patches}
\label{subsubsec_method:Removing Boundry Craters from Patches}

In our proposed system, the complete mosaic is divided into overlapping patches.
Therefore the craters occurring at the boundary of the patches are split into multiple parts, and each of the parts is assigned to corresponding patches. 
Although our system is trained to detect complete craters, the model is likely to detect partial craters as well.
Such partial craters, spread across multiple patches, are hereafter referred to as boundary craters, and their detection is undesirable. 
The output of Mask R-CNN contains information of bounding boxes and has no information related to the completeness of craters. 
Therefore, boundary craters are decided by the distance (denoted by $m$) of bounding boxes from the first and last rows and columns of the corresponding patch. We are removing boundary craters to improve precision. 
This section's experimental results are explained in Section~\ref{subsubsec_results:Removing Boundary Craters from Patches}.

\subsubsection{Pixel to Meter Conversion}
\label{subsubsec_methods:Pixel to Meter Conversion}
As we are considering the overlapping patches, there is a possibility that a single crater is detected in multiple patches.
The crater locations detected for each of the patches do not have information about the global location.
To remove the duplicate craters, we utilize a two-step approach.
In the first step, the local locations are mapped from the patches to the mosaic's global location using the pixel to meter coordinates conversion approach described in this section.
Finally, in the second step, we apply NMS (Section~\ref{subsubsec:NMS}) to remove the duplicate craters on the global locations.

Given a pixel coordinate ($x_{pxl}$, $y_{pxl}$, $r_{pxl}$), conversion to meter coordinate ($x_{meter}$, $y_{meter}$, $r_{meter}$) involves the following steps.
\begin{itemize}
    \item Calculate a resize factor ($\delta F$) as shown in Equation~\ref{eq:resize_fact} using the ratio of actual patch size ($ps_a$) and resized patch size ($ps_r$).
    \begin{align}
    \label{eq:resize_fact}
	\delta F & = \dfrac{ps_a}{ps_r}
    \end{align}
    \item Further, utilize the top right side longitude ($x_{min}$), and latitude ($y_{max}$) information (in meters) alongside the resolution ($S$) in meters/pixel.
    Finally, conversion from pixel coordinate ($x_{pxl}$, $y_{pxl}$, $r_{pxl}$) to meter coordinate ($x_{meter}$, $y_{meter}$, $r_{meter}$) is done using  Equations~\ref{eq:px_to_mtr},~\ref{eq:px_to_mtr1}, and~\ref{eq:px_to_mtr2}.   
    \begin{align}
    \label{eq:px_to_mtr}
	x_{meter} & = x_{min} + x_{pxl} \times S \times \delta F \\
	\label{eq:px_to_mtr1}
	y_{meter} & = y_{max} - y_{pxl} \times S \times \delta F \\
	\label{eq:px_to_mtr2}
	r_{meter} & = r_{pxl} \times S \times \delta F
\end{align}
\end{itemize}

In our case, $S=100$ meter/pixel, $ps_a\in\{4096,1024\}$ pixels, and $ps_r=512$ pixels.

\subsubsection{Non-Maximum Suppression (NMS)}
\label{subsubsec:NMS}
In conventional object detection, NMS is a popular approach to eliminate multiple bounding boxes for a single object~\cite{redmon2016you, girshick2014rich}. 
It is a class-specific approach to select the bounding box with the highest probability and suppress all other bounding boxes, which has an intersection over union (IOU) greater than a certain IOU threshold. 
IOU between two bounding boxes $b_i$ and $b_j$ is defined in Equation~\ref{eq:iou}.
\begin{equation}
\label{eq:iou}
IOU(b_i, b_j) = \dfrac{area(b_i \cap b_j)}{area(b_i \cup b_j) } 
\end{equation} 
Suppose there are $N$  bounding boxes, each with a detection probability $p_i$ associated with it ($i \in \{1,2,...,N\}$). 
Select the bounding box $b_h$ with the highest probability $p_h$. Compute the IOU of $b_h$ with $b_i$ where $i \in \{1,2,...,N\} \setminus \{h\}$. 
All the bounding boxes, $b_i$, with $IOU\geq \delta$ are suppressed. Here $\delta \in [0,1]$, is the range of IOU threshold. 
This entire process is repeated until all the bounding boxes are either selected or suppressed. 

\subsection{Evaluation Metrics for Crater Detection Task}

To evaluate the algorithm's effectiveness in detecting crater bounding boxes, we compute the precision, recall, and $F_1$-scores as described in this section. For a set of boxes, $ B_o= \{ b_i \mid 1\leq i \leq {N}\}$ where $ N $ (cardinality of the set $ B_o$, denoted as $\left|B_o\right|$) is the total number of detected boxes after post-processing, and a set of ground truth (craters in catalog) boxes $B_g = \{ b_j \mid 1 \leq j \leq M \}$ where $M$ is the total number of ground truth boxes. 
Evaluation metrics precision, recall and $F_1$-score are estimated for a selected threshold, u. 
Precision $P_u$ is defined as, 

\begin{equation}
\label{eq:precision}
P_u = \dfrac{ TP_u}{TP_u + FP_u} 
\end{equation}
where, $$ TP_u= \left|\{b_i |b_i\in B_o, IOU(b_i, B_g)\geq u\}\right|, $$
$$IOU(b_i, B_g) = \max_{b_m \in B_g}{}\{ IOU(b_i,b_m) \}$$
$$ FP_u= \left|B_o\right|-TP_u $$

Similarly, recall is defined as,

\begin{equation}
R_u = \dfrac{ TP_u}{TP_u + FN_u}
\end{equation}
where, $FN_u= \left|B_g\right|-TP_u $

$F_{1u}$-score is a harmonic mean of precision and recall and it is defined as,

\begin{equation}
F_{1u} = 2 \frac{P_u R_u}{P_u + R_u}
\end{equation}

Precision gives a measure of our predictions' correctness, and recall gives a measure of how well our predictions match with the catalog. 
$F_1$-score makes a balance between precision and recall.
Similar to~\cite{emami2019crater}, the value of IOU threshold ($u$) is chosen to be $0.3$ to emphasizing the recall.

\section{Experimental Results}
\label{sec:experimental results}

For this paper, we used  Keras with TensorFlow backend implementation of~\cite{matterport_maskrcnn_2017} of Mask R-CNN. 
We tuned the various hyperparameters of Mask R-CNN to make it suitable for crater detection. 
We initialized the network parameters with a pre-trained model on imagenet dataset~\cite{deng2009imagenet}.  

For the Mask R-CNN input, data augmentation is also used for the following reasons. 
The amount of shadow and bright parts in optical images of craters depends on sun angle. 
Thus, a system trained on craters from one region of a planetary surface might not generalize well in all other regions of the surface. 
In the proposed system, augmented data is generated by rotating the images by 90, 180, or 270 degrees, modifying brightness from 80\% to 150\% from the original value, applying Gaussian blur with sigma value 0 to 5, and horizontally and vertically flipping around the center. 

To focus on a particular size of craters and handle a large number of craters, we train models separately for diameter $1-2,\; 2-3,\;3-5,\;5-20,\;\text{and greater than }20$ km.

\subsection{Dataset}
\label{subsec:dataset_gen}

Optical images (monochorme, 643 nm) collected by the lunar reconnaissance orbiter camera's (LROC) wide-angle camera (WAC) on the lunar reconnaissance orbiter (LRO) (\cite{bib:robinson2010lunar}) and archived at USGS~\footnote{\url{https://astrogeology.usgs.gov/search/map/Moon/LRO/LROC\_WAC/Lunar\_LRO\_LROC-WAC\_Mosaic\_global\_100m\_June2013}} are used in this work. 
The DEM data used in the current work is constructed using the information from lunar orbiter laser altimeter (LOLA) onboard the LRO and SELENE terrain camera (TC) (\cite{bib:barker2016new}) and is archived at USGS~\footnote{\url{https://astrogeology.usgs.gov/search/map/Moon/LRO/LOLA/Lunar\_LRO\_LOLAKaguya\_DEMmerge\_60N60S\_512ppd}}. 
The resolutions for optical and DEM mosaic are 100 m/pixel and 59 m/pixel, respectively. 
The DEM data is resampled to 100 m/pixel using the ArcGIS resample tool to match their optical image's resolution. 
This step is needed as optical image, DEM, and slope raster are used simultaneously in our work.  
Slope raster is derived from DEM using the ArcGIS slope tool.

\begin{figure*}[ht!]
\centering
	\includegraphics[width=\textwidth,trim=2cm 10.8cm 2cm 10.8cm, clip]{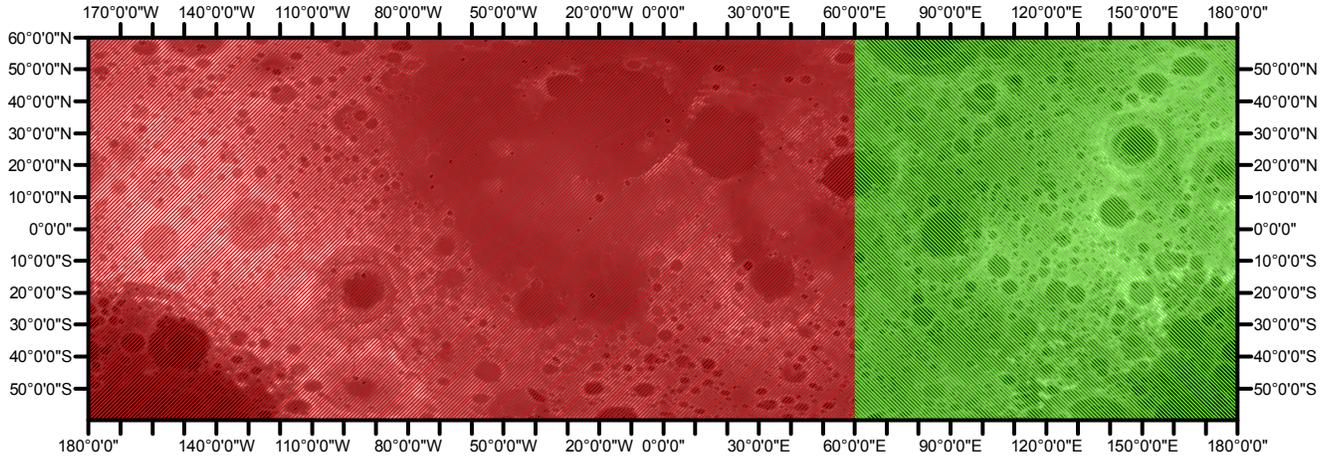}
	\caption{Train and test division from the total area chosen. The training area shaded in red, which spans $ \pm 60^{\circ}$ in latitude and $ -180^{\circ}$ to $60^{\circ}$ in longitude. The area shaded in green is used for testing purpose and spans $ \pm 60^{\circ}$ in latitude and $ 60^{\circ}$ to $180^{\circ}$ in longitude}
	\label{fig:LRO_Train_test_area}
\end{figure*}

Figure~\ref{fig:LRO_Train_test_area} shows the training area shaded in red, which spans $ \pm 60^{\circ}$ in latitude and $ -180^{\circ}$ to $60^{\circ}$ in longitude. 
It has a total size of 36388$\times$72776 pixels. 
The area shaded in green is used for testing purpose and spans $ \pm 60^{\circ}$ in latitude and $ 60^{\circ}$ to $180^{\circ}$ in longitude. 
It has a total size of 36388$\times$36388 pixels.

The primary objective of this paper is to detect maximum possible craters with a broader diameter range (craters of various sizes).
Several combinations of the input settings, such as original patch size, overlap percentage, and the patch size fed to the Mask R-CNN, the smallest crater detectable by the Mask R-CNN, need to be fixed. In this work, overlapping patches of size $1024 \times 1024$ pixels are utilized for small size craters ($<20$ km), while for larger size craters ($\geq 20$ km), patch size of $4096 \times 4096$ pixels are used. Between two consecutive patches, $50$\% overlap is considered. Further, we resized to $512 \times 512$ pixels before training to the Mask R-CNN.

We have used two catalogs for the ground truth.
The first catalog, termed as Head-LROC in this paper, is obtained by combining the catalogs of~\cite{head2010global} and~\cite{povilaitis2018crater}.
The catalog of~\cite{head2010global} have marked craters diameter of size $\geq 20$ km while the other catalog~\cite{povilaitis2018crater} have marked craters diameter of size $5 - 20$ km; both these catalogs have a relatively strict manual marking of the craters. 
This combined catalog that we term as Head-LROC in this paper is also used in~\cite{silburt2019lunar}. 
The second catalog, termed as Robbins~\cite{robbins2019new} catalog, has liberal manual marking.
Hence, the Head-LROC catalog has a smaller number of craters in a given area than the Robbins catalog.
Precisely, Head-LROC has $27,931$ craters of size $\geq5$ km, while the Robbins catalog has $1.3$ million craters of size $\geq1$ km.

\subsection{Effect of Post-Processing}
In this section, we analyzed the effect of various post-processing steps used in our proposed system.
Such analysis helped in the selection of the post-processing parameter for removing the boundary craters ($m$) and NMS IOU threshold ($\delta$).
These parameters, $m$ and $\delta$ for different crater size ranges, are optimized based on the best $F_1$-score on the validation set.
As described in Section~\ref{sec:experimental results}, separate models are trained for the various crater size range.
Therefore, we optimize the value of the $m$ and $\delta$ for the various crater sizes separately.
In the following two sections, we have shown the experimental analysis to obtain the best value of $m$ and $\delta$ for crater size $\geq 5$ km in the Head-LROC catalog.
A similar procedure has been applied for deciding $m$ and $\delta$ for all the other crater size ranges.
Note that the value of $m$ and $\delta$ are optimized simultaneously via grid search over $m\in\{0,1,5,10\}$ and $\delta \in \{0.1,0.2,0.3,0.4,0.5\}$ and without NMS scenario for best $F_1$-score on the validation set.
Hence, we have $4\times 6$ joint grid search of $m$, and $\delta$ (without NMS scenario included) for the best $F_1$-score.
For the better understanding of the parameters $m$ and $\delta$, we have only shown $4\times 1$ grid search of $m$ for the best $\delta$ value, and $1\times 6$ grid search of $\delta$ for the best $m$ value.
Finally, these jointly optimized values of $m$ and $\delta$ are utilized for the performance evaluation on the test set.

\subsubsection{Removing Boundary Craters from Patches}
\label{subsubsec_results:Removing Boundary Craters from Patches}

As we have mentioned in Section~\ref{subsubsec_method:Removing Boundry Craters from Patches}, the model is likely to detect the partial craters around the boundary, which is undesirable.
The parameter $m$ is responsible for removing the boundary crater.
As described in Section~\ref{subsubsec_method:Removing Boundry Craters from Patches}, all the detected craters within $m$ pixels of the patch boundaries are removed.
For example, if $m=5$, then any detected crater which is $5$ or fewer pixels near the patch's boundary is removed.
Table~\ref{tab:Rmv_bdry} shows that increasing the value of $m$, as expected, increases the precision with a slight decrease in recall. 
However, the F$_1$ score, which is the harmonic mean of precision and recall, increases with the value of $m$.
Hence, the value of $m$ is chosen to be 10 for all the craters of size $\geq 5$ km.
Here, Table~\ref{tab:Rmv_bdry}, represents the variation of $m$ for a fix value of $\delta=0.2$.

\begin{table}[ht!]
	\centering
	\caption{Effect of removing boundary craters in Head-LROC ($\delta:0.2$) }\label{tab:Rmv_bdry}
	\begin{tabular}{p{0.06\textwidth} p{0.09\textwidth} p{0.09\textwidth}p{0.09\textwidth}} 
		\hline
		$m$ & Precision (\%) & Recall (\%) & F$_1$-score (\%) \\ [0.5ex] 
		\hline
	    0 & 58.16 & 95.83 & 72.39 \\ 
		1 & 62.21 & 95.71 & 75.40 \\ 
		5 & 64.42 & 94.64 & 76.66 \\
		10 & 65.04 & 94.39 & 77.01 \\
		\hline
	\end{tabular}
\end{table}

\subsubsection{Effect of NMS threshold ($\delta$)}
\label{subsubsec:NMS_exp}

As described in Section~\ref{subsubsec:NMS}, NMS is used to remove the duplicate craters.
NMS threshold $\delta$ is the parameter responsible for removing the extent of duplicate craters.
Based on the definition NMS threshold $\delta$, it is expected that the higher value of $\delta$ will result in lower precision but at the same time higher recall.
This trend can also be observed in Table~\ref{tab:Head-LROC_NMS}.
The last row of the Table~\ref{tab:Head-LROC_NMS}, represents the performances for the without NMS scenario.
The importance of NMS can be seen in Table~\ref{tab:Head-LROC_NMS}, where the precision value is drastically lower for without NMS case.
The last row confers the performance without NMS. 
It can be seen that the $F_1$-score improved drastically using NMS.
Finally, the value of $\delta$ is chosen to be 0.2 for all the craters of size $\geq 5$ km.
Here, Table~\ref{tab:Head-LROC_NMS}, represents the variation of $\delta$ for a fix value of $m=10$.

To reemphasize, a similar $4\times6$ grid search is done for all the other crater size ranges, and the corresponding best $m$, and $\delta$ values are summarized in Table~\ref{tab:m_delta}.

\begin{table}[ht!]
	\centering
	\caption{Post Processing using NMS in Head-LROC ($m:10$)}\label{tab:Head-LROC_NMS}
	\begin{tabular}{p{0.09\textwidth} p{0.09\textwidth} p{0.09\textwidth}p{0.09\textwidth} } 
		\hline
		$\delta$ & Precision (\%) & Recall (\%) & F$_1$-score (\%) \\ [0.5ex] 
		\hline
		0.1 & 65.43 & 92.65 & 76.69 \\ 
		0.2 & 65.04 & 94.39 & 77.01 \\
		0.3 & 64.40 & 94.81 & 76.71 \\
		0.4 & 62.84 & 95.11 & 75.68 \\
		0.5 & 60.64  & 95.19 & 74.09 \\
		Without NMS & 11.45 & 95.37 & 20.45 \\
		\hline
	\end{tabular}
\end{table}

\begin{table}[ht!]
	\centering
	\caption{$m$ and $\delta$ value for different crater size range}\label{tab:m_delta}
	\begin{tabular}{p{0.12\textwidth}  p{0.08\textwidth}  p{0.08\textwidth} } 
		\hline
		Size Range & $m$ & $\delta$  \\ [0.5ex] 
		\hline
		$\geq 5$ km & 10 & 0.2  \\ 
		$3-5$ km & 10 & 0.3  \\
		$2-3$ km & 10 & 0.2  \\
		$1-2$ km & 1 & 0.2  \\
		\hline
	\end{tabular}
\end{table}

\subsection{Crater detection on the Lunar Surface}
\label{subsec:Crater detection on the Lunar Surface}


For the crater detection on the lunar surface, as mentioned in Section~\ref{subsec:dataset_gen}, the training area ranges from $ \pm 60^{\circ}$ in latitude and $ -180^{\circ}$ to $60^{\circ}$ in longitude with a total size of 36388$\times$72776 pixels. 
Furthermore, the test area ranges $ \pm 60^{\circ}$ in latitude and $ 60^{\circ}$ to $180^{\circ}$ in longitude with a total size of 36388$\times$36388 pixels.
These train and test regions are highlighted in Figure~\ref{fig:LRO_Train_test_area} in red and green colors, respectively.

Column 2 of the Table~\ref{tab:Total_craters}, exhibit the total number of craters of various sizes listed in the Head-LROC and Robbins catalogs of the selected region.
We further show in the Table~\ref{tab:Total_craters} the number of training and testing craters based on the different crater size range.
The training region is further divided into training and validation. 
The total number of training and validation patches are $7755$, $1974$, respectively.
From the testing region, the total number of generated patches is $5041$.

\begin{table}[ht!]
	\centering
	\caption{Distribution of Craters}\label{tab:Total_craters}
	\begin{tabular}{p{0.15\textwidth} p{0.08\textwidth}p{0.08\textwidth}p{0.08\textwidth} } 
		\hline
		Catalogs & Total  & Training  & Testing  \\ [0.5ex] 
		\hline
		Head-LROC & 24,164 & 14,602 & 9,562 \\ 
		Robbins ($\geq 5$) km & 69,097 & 38,837 & 30,260 \\
		Robbins ($3-5$) km & 101,882 & 53,015  & 48,867 \\
		Robbins ($2-3$) km &  182,537 & 95,233 & 87,304  \\
		Robbins ($1-2$) km & 679,485 & 398,404 & 281,081 \\
		\hline
	\end{tabular}
\end{table}

We tuned the network hyper-parameters such as optimizer, learning-rate, backbone architecture on the Head-LROC catalog.
Based on the experiments, SGD optimizer with momentum $0.9$, learning rate $0.001$, and ResNet-50 with FPN are utilized for the crater detection problem addressed in this paper.
Other tuned Mask R-CNN hyper-parameters for different models based on the crater size and catalog used, are given in Appendix (Table~\ref{tab:parameters_greater_5} and~\ref{tab:parameters_lesser_5}).

The nomenclature use for different models is <catalog name>\_<lowest crater size>\_<highest crater size>,
e.g., model Rob\_5\_20 is evaluated using Robbins catalog with crater diameter size range from 5-20 km. 
If the nomenclature is, e.g., Rob\_5, it indicates that evaluation is done on crater diameter size range $\geq$ 5km.

\subsection{Detection using Models Trained with Head-LROC Catalog}
\label{subsec:hl_trained}

To recall, Head-LROC catalog is a combination of two catalogs, one with  marked  crater size of $5 - 20$ km~\cite{povilaitis2018crater} and the other with crater size of $\geq 20$ km~\cite{head2010global}.
For the detection of the entire crater range of $\geq 5$ km, we train two separate models, one for the crater range $5 - 20$ km (HL\_5\_20) and the other for the crater range $\geq 20$ km (HL\_20).
Furthermore, we combine the detection results of these two models (HL\_5\_20 and HL\_20) to get the detection results for the entire crater range (HL\_5) on the Head-LROC catalog. 
The performance of both individual models and their combined one for the evaluation on the test regions with the Head-LROC and Robbins catalog are shown in Table~\ref{tab:hl_trained_hl_test} and Table~\ref{tab:hl_trained_rb_test}, respectively.

We compare our detection results with~\cite{silburt2019lunar} (referred to as Deepmoon hereafter). 
We have chosen the same DEM mosaic (59 meters/pixel) and Head-LROC catalog used in Deepmoon~\cite{silburt2019lunar}. For fair comparative analysis with Deepmoon~\cite{silburt2019lunar}, we have considered the same test region on the lunar surface.

During testing, we do not count the smaller detected craters ($< 5$ km) as false positive or true positive for the calculation of performance matrices with the Head-LROC catalog.
The strategy of removing the smaller detected craters is needed as the Head-LROC catalog has only the marking of craters with a size of $\geq 5$ km. 
A similar approach of removing smaller detected craters has also been employed in~\cite{emami2019crater} during the performance evaluation.
For a fair comparison, detected craters with a size smaller than 5 km are not counted as true positive or false positive in ours as well as Deepmoon~\cite{silburt2019lunar} work. 
For calculating the metrics (precision, recall anf F$_1$ score) in Deepmoon~\cite{silburt2019lunar} we use their post-processed output available at the link~\footnote{\url{https://doi.org/10.5281/zenodo.1133969}}. 

As described in Section~\ref{subsec:dataset_gen}, the Head-LROC catalog is incomplete and has missed many craters~\cite{silburt2019lunar}. 
Consequently, the detected craters that are not in the Head-LROC catalog will penalize the precision value. 
Hence, the F$_1$ score, which is the harmonic mean of precision and recall, is not a reliable score for the detection performance in the Head-LROC catalog; precision and recall should be analyzed separately in the Head-LROC catalog.
 
Table~\ref{tab:hl_trained_hl_test}, presents the results with models trained and tested with Head-LROC catalog.
The recall value for all crater size range is consistently on the higher side, which implies that the majority of the craters of the Head-LROC catalog are detected as the recall is 94.28\%  for the proposed method as shown in Table~\ref{tab:hl_trained_hl_test}. 
With comparison to Deepmoon~\cite{silburt2019lunar}, the improvement in the recall is 8.84\%, 2.44\%, and 6.25\% for crater size range 5-20, $\geq$20, and $\geq$5 km, respectively. 

The rationale behind the lower precision value for the proposed method and Deepmoon~\cite{silburt2019lunar} with Head-LROC catalog in Table~\ref{tab:hl_trained_hl_test} can be argued as follows.
Head-LROC has many missed craters; lower precision can be either due to newly detected craters (not marked in Head-LROC catalog) or false positives. 
This ambiguity can not be resolved with the Head-LROC catalog, and in this scenario, the Robbins catalog helps to resolve the ambiguity to an extent.
Hence, for this purpose, we use the Robbins catalog that has relatively liberal crater markings and contains the highest number of craters on the lunar surface.
Therefore if the detected craters are present in the Robbins catalog, those carters can be considered as true craters. 
Table~\ref{tab:hl_trained_rb_test} shows such detection results with the Head-LROC catalog trained model evaluated on the test regions using the Robbins catalog.
High precision values (93.21\% for crater diameter size $\geq 5$ km (HL\_5)) in Table~\ref{tab:hl_trained_rb_test} confirms that most of the newly detected craters are indeed true craters.

Another key observation in Table~\ref{tab:hl_trained_rb_test} is the drastically reduced recall values when evaluated with the Robbins catalog for Head-LROC trained model.
This reduction in the recall is expected as the model trained with Head-LROC catalog in which marking is done strictly and consider highly certain craters whereas in Robbins marking is done liberally.
Therefore, the Robbins catalog likely contains craters that are either false positive or highly degraded, which is also mentioned in~\cite{ali2020automated}. 
So Head-LROC trained model is not likely to detect these types of features of craters.
Still, compared to Deepmoon~\cite{silburt2019lunar}, the proposed method detects 4.11\% more craters in $\geq$5 km range.

\begin{table}[ht!]
	\centering
	\caption{ Performance of Head-LROC trained model in Head-LROC catalog}
	\label{tab:hl_trained_hl_test}
	\begin{tabular}{p{0.176\textwidth} p{0.06\textwidth} p{0.06\textwidth}p{0.06\textwidth} } 
		\hline
		Model Name   & Precision (\%)  & Recall (\%) & F$_1$-score (\%) \\  
		\hline
		HL\_5\_20 (Ours)  & 54.67 & \textbf{96.15} & 69.70 \\
		HL\_5\_20 (Deepmoon~\cite{silburt2019lunar})  & \textbf{61.67} & 87.31 & \textbf{72.29} \\ 
		\hline
		HL\_20 (Ours)  & \textbf{78.61} & \textbf{86.48} & \textbf{82.36} \\
        HL\_20 (Deepmoon~\cite{silburt2019lunar})  & 66.56 & 84.04 & 74.29 \\ 
		\hline
        HL\_5 (Ours) & 61.80 & \textbf{94.28} & \textbf{74.66}\\
		HL\_5 (Deepmoon~\cite{silburt2019lunar})   & \textbf{63.47} & 88.03 & 73.76 \\ 
		\hline		
	\end{tabular}
\end{table}

\begin{table}[ht!]
	\centering
	\caption{ Performance of Head-LROC trained model in Robbins catalog}
	\label{tab:hl_trained_rb_test}
	\begin{tabular}{p{0.176\textwidth} p{0.07\textwidth} p{0.07\textwidth}p{0.07\textwidth}} 
		\hline
		Model Name   & Precision (\%)  & Recall (\%) & F$_1$-score (\%) \\  
		\hline
		HL\_5\_20 (Ours) & 90.86 & \textbf{42.59} & \textbf{57.99} \\
		HL\_5\_20 (Deepmoon~\cite{silburt2019lunar})  & \textbf{93.56} & 37.10 & 53.13 \\ 
		\hline
		HL\_20 (Ours)  & \textbf{86.96} & \textbf{67.77} & \textbf{76.17} \\
		HL\_20 (Deepmoon~\cite{silburt2019lunar})  & 80.09 & 71.63 & 75.62 \\ 
		\hline
        HL\_5 (Ours) & \textbf{93.21} & \textbf{44.93} & \textbf{60.63}\\
		HL\_5 (Deepmoon~\cite{silburt2019lunar})   & 93.14 & 40.82 & 56.76\\ 
		\hline		
	\end{tabular}
\end{table}

\subsection{Detection using Models Trained with Robbins Catalog}
\label{subsec:rb_trained}

Publicly available Robbins catalog have marked craters of size $\geq$1 km.
To the best of our knowledge, the Robbins catalog has the largest number of marked craters.
To handle these large numbers of craters and emphasize a particular size range of craters, we train a separate model for each of the crater size range 1-2 km, 2-3 km, 3-5 km, 5-20 km, and $\geq$ 20 km.
Detection results of models for 5-20 km (Rob\_5\_20) and $\geq$ 20 km (Rob\_20) are combined to get the detection results for $\geq$ 5 km (Rob\_5).  
Table~\ref{tab:rb_trained_rb_test} represents the performance of Robbins trained models in Robbins catalog.

Consider the recall value for crater size $\geq 5$ km (Rob\_5), that is $75.38\%$, however when evaluated with Head-LROC catalog with Head-LROC trained model (Table~\ref{tab:hl_trained_hl_test}), the recall value is 94.28\%.
So a decrease in recall value for the Robbins trained model when evaluated with the Robbins catalog can be due to the following two reasons.
(1) The lower resolution images (100 meters/pixel) were used for training the model compared to the resolution of the Robbins catalog used for marking the lunar craters. 
(2) The Robbins catalog has liberal marking. 
Hence there is a possibility of it having false craters, similarly argued in~\cite{ali2020automated}.
These false craters are likely to confuse the model between the crater or non-crater features. 
An empirical explanation of the above argument is also given in Section~\ref{sec:Problem in Robbins Catalog}.
Therefore, before modifying Mask R-CNN's architecture for better performance for crater detection in the Robbins catalog, first, we need to find a better way of utilizing the Robbins catalog, which is part of the future work.

Further, it is observed in the Table~\ref{tab:rb_trained_rb_test} that the performance in terms of recall and F$_1$-score degrades for small size craters.
This decrease in the performance is expected as the Robbins catalog utilized high-resolution images for crater marking compared to the image resolution (100 meters/pixel) used for training the model.
It is likely that some of the craters, which are visible in the high-resolution image, may not be visible in the lower resolution image. 
It is also difficult to visually distinguish between craters and non-craters in lower-sized craters, say the craters of size range 1-2 km. 
The above argument suggests that the higher resolution training images will result in improved crater detection performance, especially for smaller size craters.

Furthermore, in Table~\ref{tab:rb_trained_hl_test}, the performance of the Robbins trained model is evaluated with the Head-LROC catalog.
Our motive for such cross-catalog detection is to find out if 
the Robbins catalog trained model can detect most of the craters present in the Head-LROC catalog. 
Recall value of 95.78\% reflects that one can detect the majority of the crater, albeit with the lower precision (31.48\%).

\subsection{Crater Localization Performance}

Crater localization here refers to the IOU between the detected crater and crater in the catalog.
In Table~\ref{tab:localization}, localization performance in terms of the mean and standard deviation of percentage IOU between the detected crater and crater in the catalog. 
For large crater sizes, the localization performance with the Robbins catalog is on the higher side compared to smaller crater sizes. These results are the consequence of better crater detection performance (Table~\ref{tab:rb_trained_rb_test}) in terms of precision, recall, and F$_1$-score for large size craters.

\begin{table}[!htb]
	\centering
	\caption{Performance of Robbins trained Model in Robbins catalog}\label{tab:rb_trained_rb_test}
	\begin{tabular}{p{0.16\textwidth} p{0.06\textwidth} p{0.06\textwidth}p{0.06\textwidth}  } 
		\hline
		Dataset & Precision (\%) & Recall (\%) & F$_1$-score (\%) \\ [0.5ex] 
		\hline
		Rob\_1\_2 & 78.47 & 54.41 & 64.26 \\
		Rob\_2\_3 & 65.65 & 62.67 & 64.13  \\
		Rob\_3\_5 & 61.02 & 74.67  & 67.16 \\
		Rob\_5\_20  & 74.15 & 75.67 & 74.91 \\
		Rob\_20  & 83.76 & 78.25 & 80.91 \\ 
		Rob\_5  & 78.40 & 75.38 & 76.86 \\
		\hline
	\end{tabular}
\end{table}

\begin{table}[ht!]
	\centering
	\caption{Performance of Robbins trained Model in Head-LROC catalog}\label{tab:rb_trained_hl_test}
	\begin{tabular}{p{0.16\textwidth} p{0.06\textwidth} p{0.06\textwidth}p{0.06\textwidth}  } 
		\hline
		Dataset & Precision (\%) & Recall (\%) & F$_1$-score (\%) \\ [0.5ex] 
		\hline
		Rob\_5\_20  & 26.75 & 97.47 & 41.98 \\ 
		Rob\_20  & 67.31 & 88.76 & 76.56 \\ 
		Rob\_5  & 31.48 & 95.78 & 47.39 \\
		\hline
	\end{tabular}
\end{table}

\begin{table}[ht!]
	\centering
	\caption{Localization performance of the proposed method}\label{tab:localization}
	\begin{tabular}{p{0.12\textwidth} p{0.09\textwidth} p{0.09\textwidth} } 
		\hline
		Dataset & Mean (\%) & Std (\%) \\ [0.5ex] 
		\hline
		HL\_5 & 81.67 & 09.31\\ 
		Rob\_5 & 83.03 & 11.35\\
		Rob\_3\_5 &  80.34 & 10.29\\
		Rob\_2\_3 &  78.03  & 09.83 \\
		Rob\_1\_2 & 68.61  & 11.65\\
		\hline
	\end{tabular}
\end{table}

\subsection{Issues in Robbins Catalog}
\label{sec:Problem in Robbins Catalog}

Robbins follows the liberal approach for manually marking craters compared to Head-LROC, which has followed a conservative approach, considering craters almost certainly. 
Consequently, the Robbins catalog has more number of manually marked craters than Head-LROC, however, it may be possible that some of the structure which is considered a crater can not be certain. 
The trained model from Head-LROC is reliable as the model is trained with a certain or almost certain crater; therefore, at detection, it will detect the crater with high certainty to be a crater. 
Whereas in Robbins, where there are likely many false positives craters that confuse the models with crater and non-crater classes, due to which detection will not be reliable and overall performance will decrease. 
Also, Robbins had used a higher resolution image to marked the crater, due to which possibly highly degraded crater $\geq$ 1 km will be detected. 
However, we trained our model on the available patches that are of comparatively lower resolution. The degraded crater is not visible due to which it severely affects the performance.

\begin{figure}[ht!]
	\includegraphics[width=\linewidth]{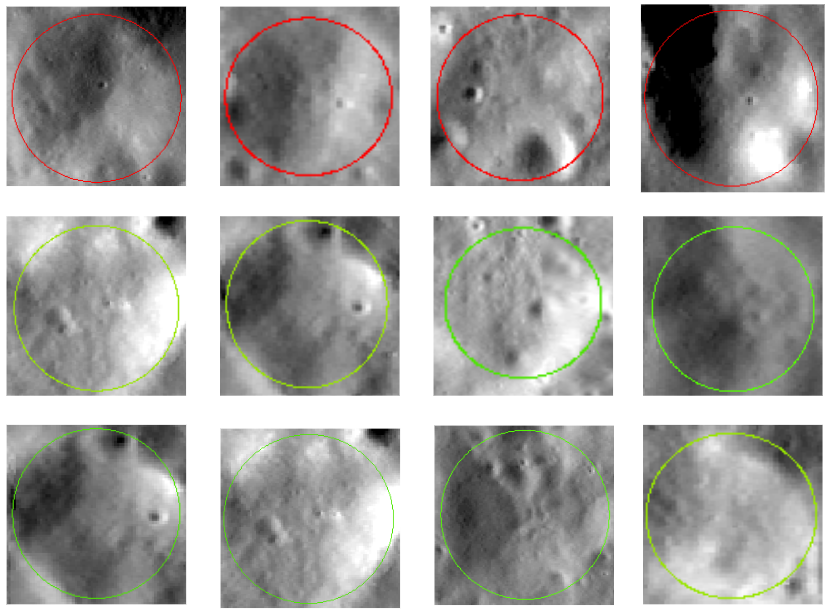}
	\caption{Visual inspection of some of the detected craters in Robbins catalog~\cite{robbins2019new}, the first row shows the craters that are detected by our model and are also present in the catalog. The last two rows show the craters, which are detected by the model but are not present in the catalog (red and green color shows the craters'  boundary)}
	\label{fig:robbins_prob_1}
\end{figure}

Figure~\ref{fig:robbins_prob_1} shows few sample craters that are either degraded or non-crater (not marked in Robbins catalog). 
The top row shows the craters that our model detected and are also present in the Robbins catalog. 
Second and third-row shows the crater detected by our model but not marked in the catalog. 
All the craters shown in the top rows and the other two rows look similar through visual inspection, but the second and third row craters are not marked in the catalog. 
Therefore there is uncertainty about whether to consider them to be a new crater detection or false positive detection.

\begin{figure}[ht!]
	\includegraphics[width=\linewidth]{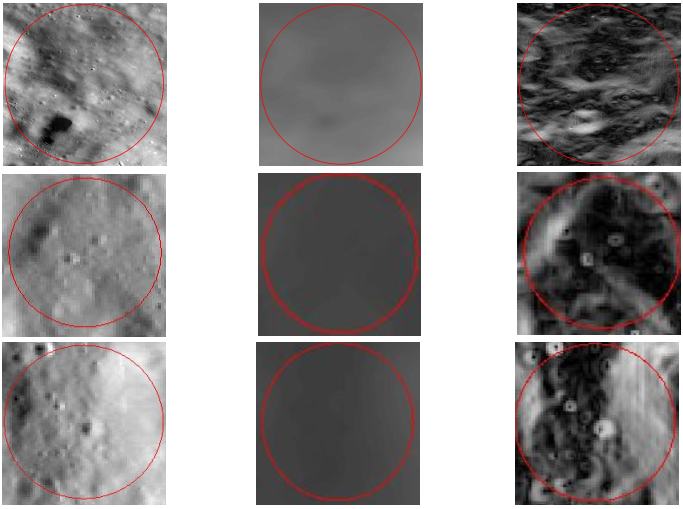}
	\caption{Visually degraded Craters in Robbins catalog~\cite{robbins2019new} in 100 m/pixel resolution images (optical image, Dem, Slope from left to right), red shows the boundary of the crater}
	\label{fig:rob_problem}
\end{figure}

\begin{figure}[ht!]
	\includegraphics[width=\linewidth]{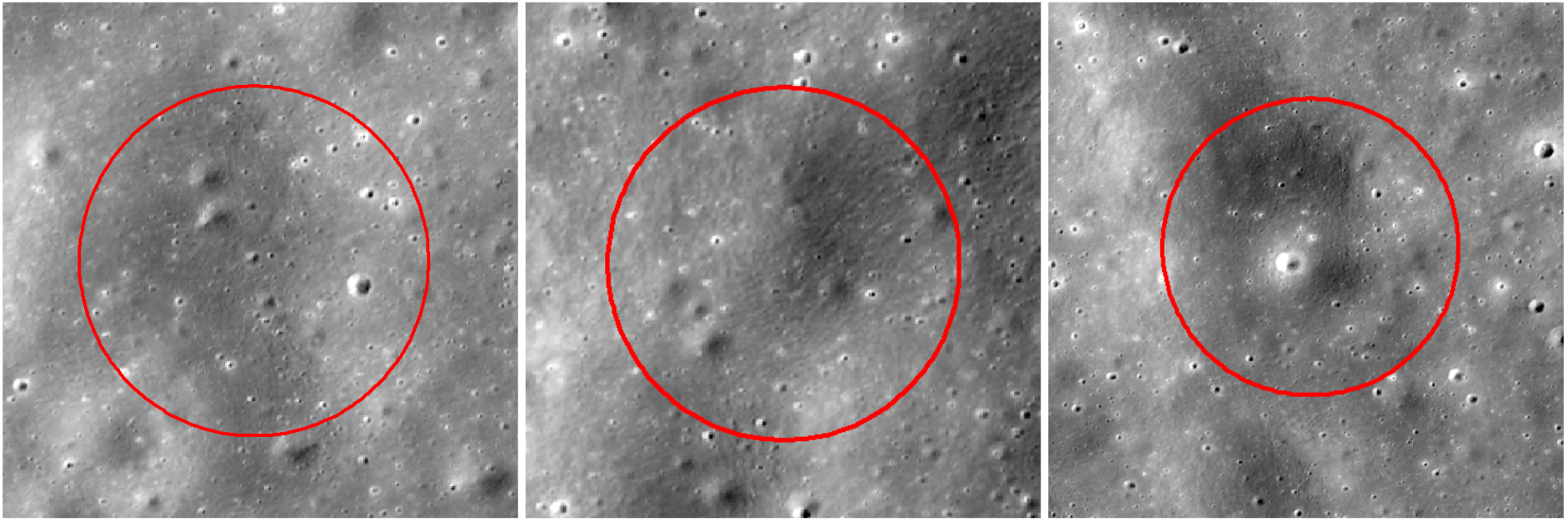}
	\caption{Visual inspection of the craters in Robbins catalog ~\cite{robbins2019new} in 5 meter/pixels resolution image of LRO NAC}
	\label{fig:robbins_high_resoltion}
\end{figure}

Figure~\ref{fig:rob_problem} shows some of the marked catalog craters from optical images, DEM, and slope, which are from 100 m/pixel resolution data. 
Through visual inspection, all seem flat surfaces in optical image, DEM, and slopes and can not be recognized as a crater. 
As Robbins catalog uses higher resolution images, there is a possibility that these craters might be visible in high-resolution images but not in the resolution we are using. 
Though in some of the cases, as shown in Figure~\ref{fig:robbins_high_resoltion}, the craters are not clearly visible even in a high-resolution LRO narrow angle camera (NAC) image~\footnote{\url{http://wms.lroc.asu.edu/lroc/rdr_product_select}} of resolution 5 m/pixel.
These issues show that we have to use Robbins' catalog more carefully.

\subsection{New Crater detection on Lunar Surface}
\label{sec:New Crater detection on Lunar Surface}

The Head-LROC catalog is conservative, but it has many missing craters. 
Robbins catalog is comparatively more complete, but it has some issues, as mentioned in the previous Section~\ref{sec:Problem in Robbins Catalog}. 
The best way to utilize both the catalog so that we detect the conservative missing crater but not false-positive craters. 
For that, we detect craters using Head-LROC trained models and consider new craters that are not present in the Head-LROC catalog and verified using Robbins catalog. 
As we are getting $93.21\%$ precision (Table~\ref{tab:hl_trained_rb_test}), which reflects that most of the new crater detected by Head-LROC trained models are reliable and can be added in Head-LROC catalogs.

\begin{figure*}[ht!]
\centering
	\includegraphics[width=0.9\linewidth]{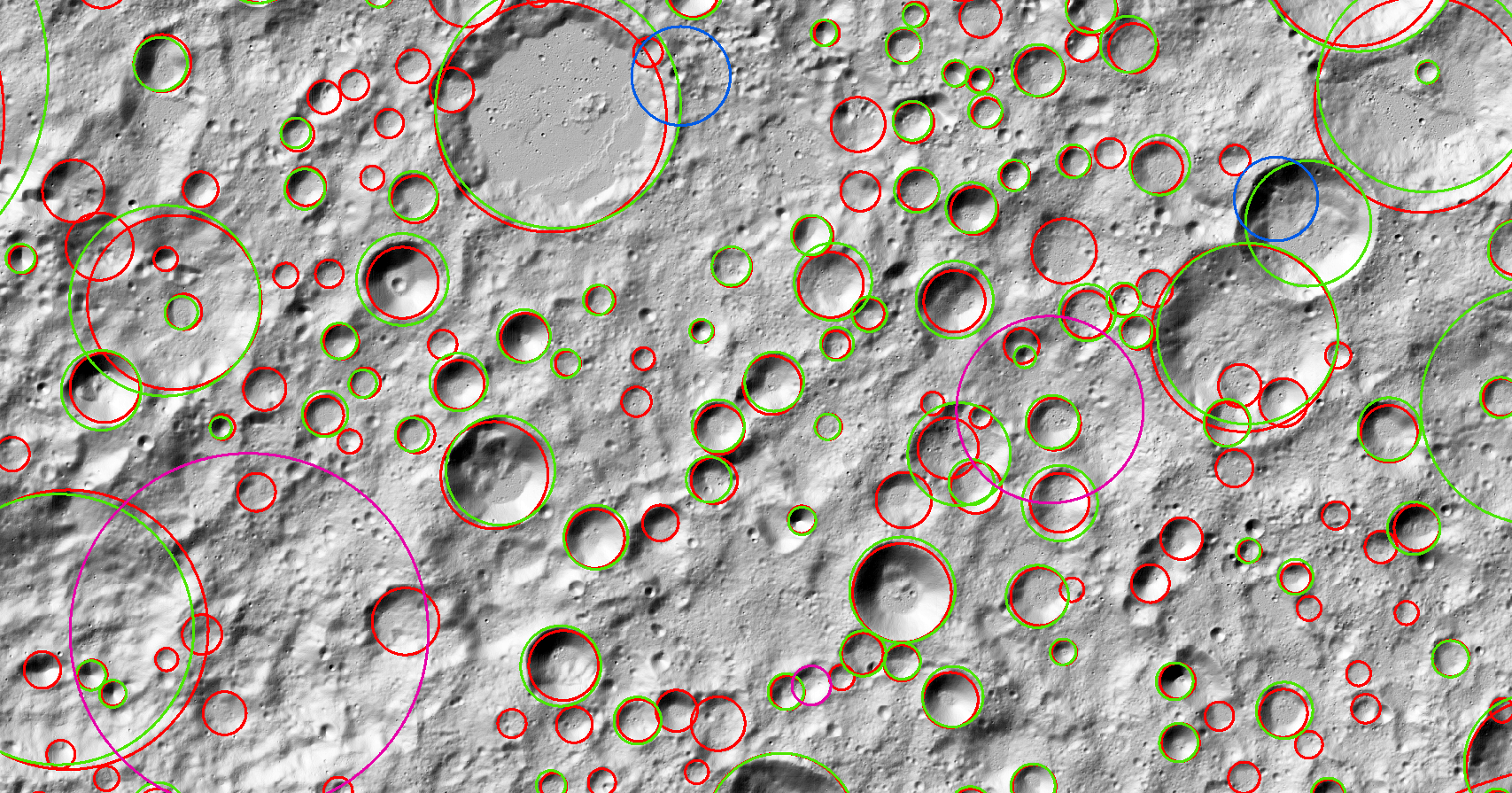}
	\caption{Visual inspection of new craters identification in some part of the test region. This shows that the most of the detected craters are true craters (Red: Detected Craters, Green: Craters in Head-LROC catalog (\cite{head2010global}, \cite{povilaitis2018crater}), Blue: False Positive, Pink: False Negative)}
	\label{fig:New Crater}
\end{figure*}

In Figure~\ref{fig:New Crater}, we have randomly chosen a subpart of the test region to inspect the detected craters on the Head-LROC trained model visually. 

Color coding of the crater boundaries, as shown in Figure~\ref{fig:New Crater} are as follows.
Craters present in the Head-LROC catalog are shown with green color,
craters detected by our model are shown in red color, false-positive craters (craters detected but not present in the Robbins catalog) are shown in blue color, while the false negative craters (craters not detected but present in the Head-LROC catalog) are shown with pink color. 
Further, the quantitative justification of the new crater's correctness in the complete test region is given in Table~\ref{tab:hl_trained_rb_test}. 
High precision ($93.21\%$) while evaluated in the Robbins catalog, shows that most of the craters detected by the Head-LROC trained model are true craters. 
Therefore, the new craters detected from the Head-LROC trained model are considered certain craters, adding the new craters to the Head-LROC catalog.

\subsection{Testing on Martian Thermal IR and DEM Data}
\label{sec:testing_on_martian_thermal_IR_and_DEM_data}

In this section, we discuss the generalization capability of our model. 
The significant difference between the Moon and Mars is the presence of the atmosphere on Mars, which would also play
a significant role in the degradation of Martian craters. 
As a result, the Martian surface shows much fewer impact craters compared to the lunar surface. 
Therefore, the performance of our algorithm on the Martian surface would be a good measure of our model's generalization capability. 
To the best of our knowledge, this is the first study that evaluates the generalization capability by using different input data types.

We utilize our model trained using the DEM, optical image, and slope (3-channel) from the lunar surface for testing on the Martian surface. In particular, we use our trained model (HL\_5\_20) for testing the performance for carters of size range 5 - 20 km on the Martian surface.
Moreover, for the Martian surface, we evaluate the generalization capability with the thermal IR images (single channel) and DEM images separately. 
Such generalization capability of the model in terms of different surface and input data types compared to training is helpful as all types of data (optical image, DEM, thermal IR) may not be available for all planetary surfaces. 

First, we tested our model on a Martian image collected by THEMIS Daytime IR images\footnote{\url{http://www.mars.asu.edu/data/thm\_dir/}}.
Specifically, we have taken thm\_dir\_N00\_000.png. 
For evaluation purposes we have considered the CSV\_Tile\_12\_0\_30\_0\_30\_px.csv from~\cite{deLatte_craterseg_2019}, which spans a longitude and latitude from $0^{\circ}$ to $30^{\circ}$.  Table~\ref{tab:Detection_Martian_th_dem} show precision, recall, and F$_1$-score with our Head-LROC trained model.

Further, we use only DEM data (one of the data types used in training) from the Martian surface to test our model's generalization capability. 
We detect the crater on the Mars HRSC MOLA Blended DEM Global 200m v2 data\footnote{\url{https://astrogeology.usgs.gov/search/map/Mars/Topography/HRSC\_MOLA\_Blend/Mars\_HRSC\_MOLA\_BlendDEM\_Global\_200mp}} (\cite{fergason2018hrsc}) and consider the same region as in the case of thermal IR data. The precision and recall (Table~\ref{tab:Detection_Martian_th_dem}) both are better as compared to testing performed with the Thermal IR Data.
This performance improvement is expected because our model was also trained using the lunar DEM data (along with optical image and slope). Therefore, detection using similar data types (DEM) as used in training is more natural to generalize than the different input data types (near IR image).

These results in Table~\ref{tab:Detection_Martian_th_dem}, show the applicability of the proposed method on the completely unknown planetary surface (Martian in our case) having a different type of input data (Thermal IR Data and DEM, in our case).
Performance can be improved further by fine-tuning the model hyper-parameters and post-processing parameters.

\begin{table}[ht!]
	\centering
	\caption{Detection Results for the model trained using the lunar surface data (HL\_5\_20) and tested on Martian Surface using THEMIS thermal IR and DEM data}\label{tab:Detection_Martian_th_dem}
	\begin{tabular}{p{0.12\textwidth} p{0.09\textwidth} p{0.09\textwidth}p{0.09\textwidth}  } 
		\hline
		Input Data type & Precision (\%) & Recall (\%) & F$_1$-score (\%) \\ [0.5ex]
		\hline
		Thermal IR data & 86.60 & 54.11  & 66.61 \\ 
		DEM data & 76.55 & 66.89  & 71.39  \\ 
		\hline
	\end{tabular}
\end{table}

\section{Conclusion and Future Work}
\label{sec:Conclusion and Future Work}

This paper proposes a deep learning-based method for crater detection on the lunar surface that is also applicable for terrain independent and data-independent crater detection.
To develop the crater detection model, we simultaneously utilized optical image, DEM, and slope data to deep learning framework to learn the crater specific features and detect them.
With our proposed method, we were able to detect the majority of the craters marked in the Head-LROC catalog and also identify the new craters. 
To check the generalization capability, we detect the crater on the Martian DEM images. 
Moreover, we showed the results on the near IR Martian images. 
Though testing with the Martian near IR images, the algorithm's performance has degraded.
Nonetheless, the results have shown that the proposed system can detect craters on a different surface and a different input data type. 
In the future, further categorization of the Robbins catalog with high-resolution images needs to be done before feeding to the deep learning architecture.

\section*{Acknowledgment}
This research is based upon work partially supported by the Indian Space Research Organisation (ISRO), Department of Space, Government of India under the Award number ISRO/SSPO/Ch-1/2016-17.  Atal Tewari was supported by TCS Research Scholarship. We are thankful to Vishal Prasad and Chennuri Prateek, students in MANAS Lab, IIT Gandhinagar, for their valuable support during the process of applying the proposed system on the Martian data. Any opinions, findings, and conclusions or recommendations expressed in this material are those of the author(s) and do not necessarily reflect the views of the funding agencies. 
We also acknowledge the use of data from the NASA's Lunar Reconnaissance Orbiter (LRO) spacecraft, which was downloaded from the archives of United States Geological Survey (USGS). Address all correspondence to Nitin Khanna at nitinkhanna@iitgn.ac.in. 

\section*{Supplementary Material}
The following link has the detection results for the Head-LROC trained model (HL\_5 in Table~\ref{tab:hl_trained_hl_test}) for the proposed method. Link:
~\url{https://docs.google.com/spreadsheets/d/1TxfaoEPPVzXqdZOgGoEDkMDq9sAGkkpQd8Wn77hBq8k/edit?usp=sharing}

\section*{Appendix}
\label{appendix}
We choose the following Mask R-CNN hyper-parameters in our work. For compactness of Table~\ref{tab:parameters_greater_5} and Table~\ref{tab:parameters_lesser_5}, RPN\_ANCHOR\_SCALES values such as ($2^{\textbf{4}}, 2^{5}, 2^{6}, 2^{7}, 2^{\textbf{8}}$) written as (\textbf{4},\textbf{8}). 

\begin{table*}[ht!]
	\centering
	\caption{Parameter values for crater size-range $\geq 5$ km}\label{tab:parameters_greater_5}
	\begin{tabular}{p{0.25\textwidth}p{0.11\textwidth}p{0.11\textwidth}p{0.11\textwidth}p{0.11\textwidth}}
		\hline
		Parameters & HL\_20  & HL\_5\_20 & Rob\_20 & Rob\_5\_20 \\  
		\hline
		TRAIN\_ROIS\_PER\_IMAGE  &  48  & 96 & 80 & 128 \\ 
		RPN\_TRAIN\_ANCHORS\_PER\_IMAGE   & 96  & 128 & 96 & 150 \\ 
		DETECTION\_NMS\_THRESHOLD   & 0.7  & 0.7 & 0.7 & 0.5 \\ 
		RPN\_NMS\_THRESHOLD    & 0.98  & 0.98 & 0.98 & 0.9\\ 
		ROI\_thresh   & 0.3  & 0.3 & 0.3 & 0.5\\ 
		POST\_NMS\_ROIS\_TRAINING  & 1000 & 1000 & 1000 & 200\\ 
		POST\_NMS\_ROIS\_INFERENCE    & 2000 & 2000 & 2000 & 400\\ 
		MAX\_GT\_INSTANCES    & 50 & 50 & 60 & 100\\ 
		DETECTION\_MAX\_INSTANCES  & 100 & 100 & 120 & 110 \\ 
		RPN\_ANCHOR\_SCALES & (4,8) & (3,7) & (4,8) & (4,8)\\
		\hline
	\end{tabular}
\end{table*}

\begin{table*}[!htb]
	\centering
	\caption{Parameter values for crater size-range $< 5$ km}\label{tab:parameters_lesser_5}
	\begin{tabular}{p{0.30\textwidth} p{0.13\textwidth} p{0.13\textwidth}p{0.13\textwidth} }
		\hline
		Parameters &  Rob\_3\_5 & Rob\_2\_3 & Rob\_1\_2 \\  
		\hline
		TRAIN\_ROIS\_PER\_IMAGE  &  128 & 200 & 650  \\ 
		RPN\_TRAIN\_ANCHORS\_PER\_IMAGE   & 256 & 256 & 700  \\ 
		DETECTION\_NMS\_THRESHOLD   & 0.7 & 0.7 & 0.7  \\ 
		RPN\_NMS\_THRESHOLD    & 0.98 & 0.98 &  0.98 \\ 
		ROI\_thresh   & 0.3 & 0.3 &  0.3 \\ 
		POST\_NMS\_ROIS\_TRAINING  & 1000 & 1000 & 1000  \\ 
		POST\_NMS\_ROIS\_INFERENCE    & 2000 & 2000 & 2000  \\ 
		MAX\_GT\_INSTANCES    & 150 & 200 & 700  \\ 
		DETECTION\_MAX\_INSTANCES  & 300 & 400 &  1400   \\ 
		RPN\_ANCHOR\_SCALES & (5,9) & (4,8) & (4,8) \\
		\hline
	\end{tabular}
\end{table*}


\end{document}